\documentclass[aps,prd,reprint,superscriptaddress,noshowpacs,preprintnumbers,longbibliography,nofootinbib,nobibnotes,floatfix]{revtex4-2}
\pdfoutput=1
\usepackage{amsmath}
\usepackage{amsfonts}
\usepackage{amssymb}
\usepackage{mathtools}
\usepackage{tabularx}
\usepackage[export]{adjustbox}
\usepackage{multirow}
\usepackage{graphicx}
\usepackage{url}
\usepackage{soul}
\usepackage{bm}
\usepackage[dvipsnames]{xcolor}
\usepackage[utf8]{inputenc}
\usepackage[colorlinks=true,breaklinks=true]{hyperref}
\hypersetup{allcolors=[rgb]{0.0 0.0 1.0},linkcolor=[rgb]{0.75 0.05 0.05}}

\hypersetup{colorlinks=true,linkcolor=blue,filecolor=magenta,citecolor=blue,urlcolor=blue}

\usepackage{orcidlink}

\begin{document}

\title[Log-likelihood-ratio statistic]{A hybrid log-likelihood-ratio statistic for single-detector SPIIR gravitational-wave searches}
\author{Weichangfeng Guo\orcidlink{0000-0002-4320-4420}}
\email{weichangfeng.guo@research.uwa.edu.au}
\author{Sunil Choudhary\orcidlink{0000-0003-0949-7298}}
\email{sunil.choudhary@uwa.edu.au}
\author{Victor Oloworaran}
\author{Linqing Wen\orcidlink{0000-0001-7987-295X}}
\email{linqing.wen@uwa.edu.au}
\author{Jayden Doan\orcidlink{0009-0005-5132-7796}}
\affiliation{Department of Physics, University of Western Australia, Crawley WA 6009, Australia}
\author{Qingyuan Liang\orcidlink{0009-0008-0278-4239}}
\affiliation{Department of Statistics, University of Auckland, 38 Princes Street, Auckland, New Zealand}
\affiliation{International Centre for Theoretical Physics Asia-Pacific (ICTP-AP), University of Chinese Academy of Sciences (UCAS), Beijing, China}
\affiliation{Taiji Laboratory for Gravitational Wave Universe (Beijing/Hangzhou),
University of Chinese Academy of Sciences (UCAS), Beijing, China}
\author{Tim Davies}
\author{Luke Davis}
\affiliation{Department of Physics, University of Western Australia, Crawley WA 6009, Australia}

\author{Irina Dvorkin\orcidlink{0000-0002-2353-9194}}
\affiliation{Institut d'Astrophysique de Paris,
CNRS, Sorbonne Universit\'e, UMR 7095, 98 bis boulevard Arago, F-75014 Paris, France}
\affiliation{Institut Universitaire de France, Minist\`ere de l’Enseignement Sup\'erieur et de la Recherche, 1 rue Descartes, 75231 Paris Cedex F-05, France}

\author{Apratim Ganguly\orcidlink{0000-0001-7394-0755}}
\affiliation{Inter-University Centre for Astronomy and Astrophysics (IUCAA), Post Bag 4, Ganeshkhind, Pune 411 007, India}

\author{Shang-Jie Jin\orcidlink{0000-0003-3697-3501}}
\affiliation{Department of Physics, University of Western Australia, Crawley WA 6009, Australia}
\affiliation{Liaoning Key Laboratory of Cosmology and Astrophysics, College of Sciences, Northeastern University, Shenyang 110819, China}
\affiliation{Research Center for Space Science, Advanced Research Laboratories, Tokyo City University, 1-28-1 Tamazutsumi, Setagaya-ku, Tokyo 158-8557, Japan}

\author{JeongCho Kim\orcidlink{0000-0003-1991-2483}}
\affiliation{National Institute for Mathematical Sciences, Daejeon 34047, Republic of Korea}

\author{Yu-Chiung Lin\orcidlink{0000-0003-4939-1404}}
\author{Ngai-Chun Lung\orcidlink{0009-0005-4174-414X}}
\affiliation{Institute of Astronomy, National Tsing Hua University,  Hsinchu 300044, Taiwan R.O.C.}
\author{Alistair McLeod\orcidlink{0000-0001-5424-8368}}
\affiliation{Department of Physics, University of Western Australia, Crawley WA 6009, Australia}
\author{Khun Sang Phukon}
\affiliation{School of Physics and Astronomy and Institute for Gravitational Wave Astronomy, University of Birmingham, Edgbaston, Birmingham, B15 2TT, United Kingdom}
\author{Andreas Wicenec\orcidlink{0000-0002-1774-5653}}
\affiliation{
International Centre for Radio Astronomy Research, University of Western Australia, Crawley WA 6009, Australia
}
\collaboration{SPIIR Collaboration}
\noaffiliation

\date{\today}

\begin{abstract}

Periods in which only one detector is operating account for a substantial fraction of gravitational-wave observing time, including 13.3\% of the third LIGO--Virgo observing run (O3) and 29.7\% of the first part of the fourth observing run (O4a). We present the first dedicated single-detector implementation of the Summed Parallel Infinite Impulse Response (SPIIR) pipeline for binary neutron star (BNS) and neutron star--black hole (NSBH) merger searches.
We construct a hybrid log-likelihood ratio that combines signal-consistency models for non-Gaussian glitches with an SNR profile-likelihood-ratio term derived under a Gaussian-noise model. The glitch model is calibrated separately by source category and trigger-rate class.
We then assign false alarm rates (FARs) using the measured background from the same detector and analysis chunk, with extrapolation of its high-ranking-statistic tail.
The FAR assignments are checked for consistency by comparing the observed cumulative background counts with those expected from the assigned FARs for the BNS and NSBH searches in the two halves of the third observing run (O3a and O3b), respectively, at LIGO Hanford (H1) and LIGO Livingston (L1).
Using O3 data and the GWTC-3 injection set, we find that, at a threshold of one per two months, the BNS search achieves sensitive distances of $136.5^{+2.9}_{-3.0}$~Mpc and $120.2^{+3.1}_{-3.3}$~Mpc for L1 and H1, compared with $110.5^{+3.2}_{-3.4}$~Mpc and $87.2^{+3.6}_{-3.9}$~Mpc for GWTC-3 GstLAL, the only published single-detector results in GWTC-3, over the same data and injection parameter ranges. The central estimates are approximately 24\% and 38\% higher, respectively. For NSBH, SPIIR reaches $164.1^{+6.7}_{-7.3}$~Mpc in L1 and $134.9^{+6.3}_{-7.0}$~Mpc in H1, compared with $165.1^{+6.7}_{-7.3}$~Mpc and $156.0^{+5.9}_{-6.4}$~Mpc for GWTC-3 GstLAL, respectively. For validation, we have performed targeted recovery of known single-detector BNS and NSBH events GW190425, GW200105\_162426, and GW230529\_181500. All three events are recovered below the detection threshold of one per two months.
\end{abstract}

\maketitle

\section{\label{sec:level1}Introduction}
Gravitational-wave (GW) astronomy has opened a new window into the Universe since the first direct detection of GW150914~\cite{gw150914} by the LIGO detectors~\cite{ligo}. Successive observing runs have expanded the operating network from the two LIGO detectors to include Virgo and KAGRA~\cite{virgo,kagra}. Improvements in detector sensitivity and search methods have led to more than two hundred compact binary coalescence (CBC) detections across successive catalogs~\cite{gwtc-1,gwtc-2,gwtc-2.1,gwtc-3,gwtc-4}, providing an increasingly large sample for studies of black-hole and neutron-star binary populations.

Despite the expansion of the detector network, substantial intervals remain in which only one detector is observing. During the first and second halves of the third observing run (O3a and O3b), one detector was observing for $15.0\%$ and $11.2\%$ of the runs, respectively, amounting to 13.3\% of the total duration. In the first part of the fourth observing run (O4a), the corresponding fraction was $29.7\%$~\cite{o3_open_data,gwtc4_intro}.

Substantial single-detector observing time motivates search methods that can assign significance to triggers without relying on detector coincidence. This is especially relevant for binary neutron star (BNS) and neutron star--black hole (NSBH) mergers: although a single detector cannot provide precise sky localization, its accurately measured merger time can still aid electromagnetic follow-up.
Several reported events had catalogued significance determined by a single detector, including the BNS merger GW190425~\cite{gw190425} and two NSBH mergers GW200105\_162426~\cite{gw200105} and GW230529\_181500~\cite{gw230529}. LIGO Livingston (L1) and Virgo (V1) were observing during GW190425 and GW200105\_162426, but the reported significance of each event was determined by the L1 trigger; GW230529\_181500 occurred when only L1 provided usable data. 

These single-detector events are also of astrophysical interest beyond their detection.
GW190425 has a source-frame total mass of about $3.4\,M_\odot$, significantly larger than those of known Galactic binary neutron-star systems~\cite{gw190425}.
Although its sky localization was poor, a retrospective search identified a potential coincidence with FRB~20190425A, detected about 2.5~hr after the merger within the gravitational-wave localization region; a possible association was proposed with a reported coincidence significance of $2.8\sigma$~\cite{gw190425_em}.
That proposal has since been examined with numerical-relativity simulations~\cite{gw190425_noem_1}, gravitational-wave parameter estimation~\cite{gw190425_noem_2}, and wide-field optical limits~\cite{gw190425_noem_3}.
GW230529\_181500 has a source-frame primary mass of $2.5$--$4.5\,M_\odot$ at 90\% credibility, in the purported lower mass gap between the most massive neutron stars and the least massive black holes, and its discovery implies a higher expected rate of NSBH mergers with electromagnetic counterparts~\cite{gw230529}.
Recovering such events with reliable significance requires a ranking statistic designed for single-detector data.

Multi-detector coincidence suppresses instrumental glitches by requiring consistent arrival times and, where applicable, phases and amplitudes. Candidate significance is quantified by the false alarm rate (FAR), typically estimated from time-shifted coincidences. A single-detector search lacks these intersite tests and time-shifted backgrounds, and its directly measured FAR range is limited by the analyzed livetime. It therefore requires a ranking statistic that separates signals from transient noise and a model for significance beyond the measured background tail.

Several GW search pipelines have developed strategies to address these challenges. The GstLAL pipeline~\cite{gstlal_prompt,gstlal_1} employs a multidimensional likelihood ratio ranking statistic~\cite{llr_1,llr_2,gstlal_ranking_improved} combining matched-filter signal-to-noise ratio (SNR) $\rho$, signal consistency tests $\chi^2$~\cite{chisq}, and data quality information from the iDQ glitch classifier~\cite{idq}. Empirical penalties are applied to single-detector triggers to suppress false alarms and prevent them from dominating the candidate list when multi-detector coincidence is absent~\cite{gstlal_single_1,gstlal_single_2}.
To suppress glitches, the MBTA pipeline~\cite{mbta_1,mbta_2} uses a reweighted SNR statistic, vetoes triggers occurring during periods of abnormally high trigger activity, and estimates the FAR of single-detector triggers by modeling the tail of the ranking statistic distribution~\cite{mbta_single_o4,mbta_single_1}. Similarly, PyCBC~\cite{pycbc_1,pycbc_3,pycbc_open,pycbc_4} assigns significance to single-detector triggers using a reweighted SNR statistic and extrapolates the FAR by an exponential fit~\cite{pycbc_single_1}.
These approaches demonstrate that astrophysical signals can be recovered in single-detector data, but their ranking and background models are specific to each pipeline.

In this work, we extend the Summed Parallel Infinite Impulse Response (SPIIR) pipeline~\cite{spiir_1,spiir_2,spiir_4,spiir_6,spiir_7,spiir_8} to searches using a single detector. The SPIIR pipeline accelerates matched filtering with infinite impulse response (IIR) filters implemented on GPUs~\cite{spiir_1,spiir_3,spiir_5}, enabling efficient low-latency searches for CBC signals. SPIIR was one of the four CBC low-latency matched-filter pipelines used during O3 and O4, together with GstLAL, MBTA, and PyCBC. Each SPIIR trigger is characterized by its peak matched-filter SNR $\rho_m$ and a reduced chi-square statistic $\chi^2_r$ that measures data consistency with the template response.

The standard SPIIR analysis coherently combines the detector responses over sky directions~\cite{spiir_7}. The candidate significance is determined from the rarity of a trigger in the background distribution from multiple detectors over the $(\rho_m,\chi^2_r)$ plane, and a $\chi^2_r$ gate suppresses transient noise~\cite{spiir_7}. However, without the help of a multi-detector coincidence test, detections with single-detector data are more vulnerable to unexpected noise transients. The noise hypothesis of a single-detector ranking statistic therefore must model the glitch-dominated background directly, and determines the final significance from the measured detector background.

We propose a hybrid likelihood ratio that takes into account non-Gaussian glitch and stationary Gaussian noise.
We assess this method using cumulative background counts and measure the sensitivity using simulated BNS and NSBH injections in the selected O3 single-detector segments. Within the GWTC-3 O3 sensitivity release used here, GstLAL is the only pipeline for which corresponding single-detector BNS and NSBH sensitivity estimates are provided~\cite{gwtc-3,sensitivity_estimate}. Relative to GstLAL, the central estimates of the BNS sensitive distance are approximately 24\% and 38\% higher in L1 and H1, respectively. For NSBH, the L1 estimates agree to within 1\%, while the H1 SPIIR estimate is approximately 14\% lower.
All three previously reported events whose catalog significance was determined from a single-detector are recovered below the adopted FAR threshold.

This paper is organized as follows. Section~\ref{sec:data} describes the data sets used in the study. Section~\ref{sec:methods} derives the log-likelihood-ratio statistic and gives the FAR extrapolation procedure. Section~\ref{sec:results} presents the FAR consistency check, the single-detector BNS and NSBH sensitivity estimates and their comparison with the GWTC-3 GstLAL results, and the recovery of known events in O3 and O4a. Section~\ref{sec:conclusion} summarizes the results and discusses future work. Appendix~\ref{app:nearby_missed} examines nearby missed injections using matched-trigger diagnostics and Q-transform maps~\cite{q_transform}, including an additional L1 NSBH data segment.

\section{\label{sec:data}Data}

In this study, we focus on applying the single-detector method on the LIGO Hanford (H1) and LIGO Livingston (L1) detector data, since the Virgo (V1) detector had significantly lower sensitivity during the third LIGO--Virgo observing run (O3).
We use C01-calibrated strain data from O3 public data~\cite{o3_open_data}. The data are downsampled to 2048\,Hz for our analysis. Following the data segmentation scheme adopted by the LVK Collaboration, the full O3 run is divided into consecutive analysis chunks. The O3 science run is separated into two segments by an intermission period: O3a covers GPS times 1238166018--1253977218 and comprises chunks~1--22; O3b covers GPS times 1256655618--1269363618 and comprises chunks~23--40.

For the O3 analyses, we construct background for each detector and analysis chunk, excluding time windows around catalogued events. Search triggers are assigned FARs using the background from the same detector and chunk. For injection studies, the background is obtained from the same data without added signals. We assess these FAR estimates using cumulative background-trigger counts from the entire O3a BNS and O3b NSBH searches.

For the GW230529\_181500 recovery test, we use O4a C00 frames and the uncleaned \texttt{L1:GDS-CALIB\_STRAIN\_AR} strain channel from O4 open data~\cite{o4_open_data}. These data are also sampled at 2048\,Hz for the SPIIR analysis. 
We apply no additional data-quality vetoes or strain gating to the analyzed strain.

\section{Methods}\label{sec:methods}

For the BNS search, the template bank covers component masses \(m_1\in[1.1,3]\,M_\odot\) and \(m_2\in[1.1,3]\,M_\odot\). For the NSBH search, the ranges are \(m_1\in[3,100]\,M_\odot\) and \(m_2\in[1.1,3]\,M_\odot\). The dimensionless spin magnitudes are limited to \(|\chi|\le0.05\) for neutron star components and \(|\chi|\le0.99\) for black hole components. The template bank is constructed using stochastic template placement~\cite{gwtc-2,gwtc-3}. Template waveforms are generated with a lower frequency cutoff of $15\,\mathrm{Hz}$, using \textsc{TaylorF2} for detector-frame chirp masses below $1.73\,M_\odot$ and \textsc{SEOBNRv4\_ROM} for chirp masses above this value. The chirp mass is defined as $\mathcal{M}=(m_1m_2)^{3/5}/(m_1+m_2)^{1/5}$.

The SPIIR search employs a template bank divided into smaller banks, each containing 1,000 waveform templates, with multiple banks processed on each computing node. Matched filtering produces a complex SNR time series, $\rho(t)$, for each template. Within each one-second window, triggers are clustered across all templates processed by a computing node, retaining only the trigger with the highest SNR. No further clustering is applied across computing nodes before FAR assignment or cumulative counting. For the retained trigger, $t_m$ is the peak time and $\rho(t)$ denotes the output of its associated template. The peak SNR is
\begin{equation}
\rho_m \equiv |\rho(t_m)|.
\label{eq:peak_snr_def}
\end{equation}

For each retained trigger, the reduced time-domain chi-square statistic $\chi_r^2$ measures the consistency of the SNR time series around $t_m$ with the template autocorrelation response~\cite{time_domain_chisq,spiir_7}. It is defined by
\begin{align}
\chi^2 &= \sum_{i=1}^{N_a} \left| \rho(t_i) - \rho(t_m) a(t_i) \right|^2, \label{eq:chi2_def_raw}\\
\chi^2_r &= \chi^2 / \nu,
\end{align}
where \(a(t_i)\) is the template autocorrelation function and \(N_a\) is the number of samples in the autocorrelation window. In Eq.~\eqref{eq:chi2_def_raw} and the equations below, \(t_i\) denotes an offset from \(t_m\), and \(\rho(t_i)\) is shorthand for \(\rho(t_m+t_i)\). The normalization constant $\nu$ is defined as
\begin{align}
\nu = 2\sum_{i=1}^{N_a}\!\left(1 - |a(t_i)|^2\right),
\end{align}
which is the expected value of $\chi^2$ at a fixed reference time under the stationary-Gaussian noise model.

\subsection{\label{sec:rs}Ranking statistic}

We construct the log-likelihood ratio from the peak SNR, $\rho_m$, and reduced chi-square statistic, $\chi_r^2$, of each retained trigger. An exact joint log-likelihood ratio would compare the selected-trigger densities under the signal hypothesis, $\mathcal{H}_1$, and the noise hypothesis, $\mathcal{H}_0$:

\begin{align}
\ln \Lambda_{\rm joint}
= \ln \frac{P(d \mid \mathcal{H}_1)}{P(d \mid \mathcal{H}_0)},
\end{align}
where $d=(\rho_m,\chi_r^2)$ contains the two observables defined above.

Applying the product rule gives
\begin{align}
\ln \Lambda_{\rm joint}
&=
\ln \frac{P(\rho_m,\chi^2_r \mid \mathcal{H}_1)}
{P(\rho_m,\chi^2_r \mid \mathcal{H}_0)} \\
&=
\ln \frac{P(\chi^2_r \mid \rho_m,\mathcal{H}_1)}
{P(\chi^2_r \mid \rho_m,\mathcal{H}_0)}
+
\ln \frac{P(\rho_m \mid \mathcal{H}_1)}
{P(\rho_m \mid \mathcal{H}_0)} .
\label{eq:lr_decomp}
\end{align}

This factorization motivates our ranking statistic, denoted by $\ln\mathcal L$. The noise background contains both Gaussian noise ($\mathcal H_G$) and instrumental glitches ($\mathcal H_g$), which can be represented as a mixture,
\begin{equation}
P(d\mid\mathcal H_0)=(1-\eta)P(d\mid\mathcal H_G)+\eta P(d\mid\mathcal H_g),
\label{eq:noise_mixture}
\end{equation}
where $\eta$ is the glitch fraction. We do not fit this full mixture. Instead, the signal-consistency term uses an effective glitch model, while the SNR term uses a stationary-Gaussian reference likelihood maximized over signal amplitude and phase. These two contributions are derived below.

\subsubsection{Distribution of \texorpdfstring{$\chi_r^2$}{chi-squared} given \texorpdfstring{$\rho_m$}{rho m}}
\label{sec:chisq_H0}

To model the distribution of $\chi_r^2$ at a given $\rho_m$, we write the matched-filter SNR time series in Eq.~\eqref{eq:chi2_def_raw} as
\begin{align}
\rho(t) = n(t) + s(t),
\end{align}
where $n(t)$ denotes the noise component of the matched-filter output and $s(t)$ denotes the signal contribution. In the glitch hypothesis, $s(t)=0$. We define the mismatch between the signal and the template normalized by the observed peak as
\begin{align}
\beta(t_i) \equiv
\frac{s(t_i)-\rho(t_m) a(t_i)}
{\rho(t_m) a(t_i)}.
\end{align}
This coefficient includes noise in $\rho(t_m)$; in the signal-dominated limit it describes waveform mismatch. Substituting this relation into
Eq.~(\ref{eq:chi2_def_raw}) yields
\begin{align}
\chi^2 =
\sum_{i=1}^{N_a}
\left| n(t_i) + \beta(t_i)\rho(t_m) a(t_i) \right|^2 .
\end{align}

The coefficient $\beta(t_i)$ is generally complex. For tractability, we represent the residual strength across the autocorrelation window by a constant real effective parameter $\bar{\beta}$. The distribution below depends on this parameter only through $|\bar{\beta}|^2$. With this approximation,
\begin{align}
\chi^2 \simeq
\sum_{i=1}^{N_a}
\left| n(t_i) + \bar{\beta}\rho(t_m) a(t_i) \right|^2 .
\label{eq:chisq_beta}
\end{align}

The residual form in Eq.~\eqref{eq:chisq_beta} motivates an effective noncentral chi-square model for the measured $\chi_r^2$. We use an effective degree-of-freedom parameter $\nu_{\rm eff}$ to describe the shape of the distribution after trigger selection. We adopt the noncentrality scale
\begin{equation}
\lambda =
|\bar{\beta}|^2\rho_m^2
\sum_{i=1}^{N_a}|a(t_i)|^2.
\end{equation}

Thresholding, peak SNR selection, trigger clustering~\cite{spiir_7}, and real detector transients alter the selected-trigger distribution. We therefore retain the analytical moment structure but replace the nominal quantities by effective parameters. To describe template-dependent differences in the selected-trigger background, we introduce an effective autocorrelation scale,
\begin{align}
A_{\rm eff}(T)
&=
\left[\sum_{i=1}^{N_a}|a(t_i)|^2\right]^\alpha,
\qquad 0\leq\alpha\leq1,
\label{eq:effective_autocorrelation_scale}\\
\lambda_{\rm eff}
&=
|\bar{\beta}|^2\rho_m^2 A_{\rm eff}(T),
\label{eq:lambda_eff}
\end{align}
where $T$ labels the template. The exponent $\alpha$ controls the contribution of the template autocorrelation to the noncentrality. We calibrate $\alpha$ jointly with $\nu_{\rm eff}$ as described in Sec.~\ref{sec:ranking_calibration}.

For the large effective degrees of freedom used here, we evaluate the \(\chi_r^2\) density with a Gaussian approximation to the corresponding noncentral chi-square expression,
\begin{align}
p_{\chi_r^2}(x \mid \rho_m,\bar{\beta})
&\simeq
\frac{1}{\sqrt{2\pi\sigma_r^2}}
\exp\left[
-\frac{(x-\mu_r)^2}{2\sigma_r^2}
\right],
\label{eq:reduced_chisq_pdf}\\
\mu_r &= 1+\frac{\lambda_{\rm eff}}{\nu_{\rm eff}},\qquad
\sigma_r^2 = \frac{2}{\nu_{\rm eff}}+\frac{4\lambda_{\rm eff}}{\nu_{\rm eff}^2},
\end{align}
where $x$ denotes the observed value of $\chi_r^2$.

We construct effective signal and glitch consistency models by marginalizing the above distribution over $\bar{\beta}$.
Formally setting \(s(t_i)=0\) in the definition of \(\beta(t_i)\) gives \(\bar{\beta}=-1\). 
For the glitch model, we adopt
\begin{align}
P(\bar{\beta}\mid\mathcal{H}_g)=\delta(\bar{\beta}+1)
\end{align}
as a phenomenological fully unsupported limit for the ranking-relevant transient population. The partial autocorrelation support of real glitches is represented by \(A_{\rm eff}(T)\). 

For the signal hypothesis, the mismatch arises from waveform differences between the signal and the best-matching template. For a signal recovered with match $M\simeq1$, the mismatch contribution to the squared matched-filter SNR residual scales approximately as $(1-M^2)\rho_m^2$~\cite{chisq}.

We accordingly take
\begin{equation}
    |\bar{\beta}| \leq \bar{\beta}_{\max}
    = \sqrt{1-M_{\min}^2}
    \approx 0.24,
\end{equation}
where $M_{\min}=0.97$ is the minimum match of the template bank.

We model $\bar{\beta}$ as a uniform random variable,
\begin{align}
P(\bar{\beta}\mid\mathcal{H}_1)=
\begin{cases}
1/2\bar{\beta}_{\max}, & -\bar{\beta}_{\max}\leq\bar{\beta}\leq\bar{\beta}_{\max},\\
0, & \text{otherwise.}
\end{cases}
\end{align}

The effective consistency models used in the ranking statistic are
\begin{align}
q_g(\chi_r^2\mid \rho_m)
&=p_{\chi_r^2}(\chi_r^2\mid\rho_m,\bar{\beta}=-1),\\
q_s(\chi_r^2\mid \rho_m)
&=\frac{1}{2\bar{\beta}_{\max}}
\int_{-\bar{\beta}_{\max}}^{\bar{\beta}_{\max}}
p_{\chi_r^2}(\chi_r^2\mid\rho_m,\bar{\beta})
\,d\bar{\beta}.
\label{eq:chisq_hypothesis_models}
\end{align}
Since the integrand is even in $\bar{\beta}$, we approximate the marginalization using an equal-weight average over 10 equally spaced points on $[0,\bar{\beta}_{\max}]$, including both endpoints.

\subsubsection{Distribution of $\rho_m$ under $\mathcal{H}_1$ and $\mathcal{H}_0$}
For the SNR contribution, we use the generalized likelihood-ratio statistic for a signal in stationary Gaussian noise,
\begin{equation}
\widehat{\ln\Lambda}_{\rm SNR}
=\max_{A,\phi}\ln
\frac{p(x,y\mid A,\phi,\mathcal H_{1,G})}
     {p(x,y\mid\mathcal H_G)},
\end{equation}
where $\mathcal H_{1,G}$ denotes a signal added to Gaussian noise.

For a unit-normalized template, we write the complex matched-filter SNR in terms of its quadratures as \begin{equation} \rho(t)=x(t)+i y(t), \end{equation} where, under the Gaussian noise hypothesis, \(x(t)\) and \(y(t)\) are independent Gaussian random variables with zero mean and unit variance.

At a fixed time sample, the joint probability density of the matched-filter quadratures under the noise hypothesis is
\begin{equation}
p(x,y\mid \mathcal H_G)
=
\frac{1}{2\pi}
\exp\left[-\frac{x^2+y^2}{2}\right].
\end{equation}

For a signal with amplitude $A$ and phase $\phi$, the mean of the matched-filter output is displaced in the complex plane~\cite{snr_llr}. The corresponding
density is
\begin{align}
p(x,y\mid A,\phi,\mathcal H_{1,G})
&=
\frac{1}{2\pi}\exp\!\left[-\frac{1}{2}Q(x,y;A,\phi)\right],
\nonumber\\
Q(x,y;A,\phi)
&=(x-A\cos\phi)^2+(y-A\sin\phi)^2 .
\end{align}

The log-likelihood-ratio at fixed \((A,\phi)\) is therefore
\begin{align}
\ln \frac{p(x,y\mid A,\phi,\mathcal H_{1,G})}
{p(x,y\mid \mathcal H_G)}
&= A(x\cos\phi+y\sin\phi)-\frac{1}{2}A^2 .
\end{align}
Maximizing over the signal phase aligns the model with the observed quadrature vector and gives
\begin{equation}
\widehat{\ln\Lambda}_{\rm SNR}(A)=A\sqrt{x^2+y^2}-\frac{1}{2}A^2 .
\end{equation}
A subsequent maximization over the non-negative amplitude gives
\begin{equation}
\widehat{A}=\sqrt{x^2+y^2}=|\rho|,
\end{equation}
and hence the log-likelihood-ratio
\begin{equation}
\widehat{\ln\Lambda}_{\rm SNR}
=
\frac{1}{2}|\rho|^2+\mathrm{const}.
\end{equation}
Evaluating this profile statistic at the retained trigger peak, we define
\begin{equation}
\ell_{\rm SNR}\equiv\frac{1}{2}\rho_m^2,
\label{eq:snr_profile_likelihood}
\end{equation}
up to an irrelevant additive constant. The noise reference is the Gaussian component of Eq.~\eqref{eq:noise_mixture}. This choice does not marginalize over an astrophysical amplitude prior or model the selected-peak noise density.

Combining Eqs.~(\ref{eq:chisq_hypothesis_models}) and~(\ref{eq:snr_profile_likelihood}), the ranking statistic used in this work is
\begin{align}
\ln\mathcal{L}
&=
\ln
\frac{
\displaystyle
\frac{1}{\bar{\beta}_{\max}}
\int_0^{\bar{\beta}_{\max}}
p_{\chi_r^2}(\chi_r^2\mid\rho_m,\bar{\beta})\,d\bar{\beta}
}{
p_{\chi_r^2}(\chi_r^2\mid\rho_m,\bar{\beta}=-1)
}
+\frac{1}{2}\rho_m^2.
\label{eq:final_ranking_statistic}
\end{align}

\subsubsection{Parameter calibration}
\label{sec:ranking_calibration}

The effective parameters $\alpha$ and $\nu_{\rm eff}$ describe the background-trigger distribution, which can vary between chunks. We calibrate the parameters separately for each search category and trigger-rate class. For detector $d$ and chunk $c$, we calculate $R_{dc}=N_{dc}/T_{dc}$, where $N_{dc}$ is the number of non-injection triggers with $\rho_m\geq8$ and $\chi_r^2>3$, and $T_{dc}$ is the analyzed livetime. These cuts select relatively loud triggers with poor template consistency, providing an empirical indicator of transient-noise activity. Both the trigger count and the livetime exclude windows of $\pm1\,\mathrm{s}$ around catalogued events. We define
\begin{equation}
Q_{dc} =
\frac{R_{dc}}{\operatorname{median}_{c}(R_{dc})},
\label{eq:chunk_rate_ratio}
\end{equation}
where the median is calculated separately for each detector and search category over O3a for BNS and O3b for NSBH. Chunks with $Q_{dc}>4$ are classified as high-rate, and the remaining chunks as nominal-rate. This classification uses the measured trigger quantities before evaluating the ranking statistic. Both classes use the same likelihood model, with different calibrated parameter values.

We use dedicated L1 injection samples for parameter calibration, drawn from chunks distinct from those used for the sensitivity estimates in Sec.~\ref{sec:sensitivity}. For each trial pair $(\alpha,\nu_{\rm eff})$, we evaluate the injection triggers and the non-injection background from the same detector and chunk with that pair. We then construct the FAR mapping as described in Sec.~\ref{sec:far} and assign FARs to the injection triggers. The NSBH sensitivity samples were examined during method development but were not used in the final parameter scan.

We scan $\alpha$ from $0.1$ to $1.0$ in steps of $0.1$, and $\nu_{\rm eff}$ in steps of $10$ over $40$--$600$. We select the parameter pair that maximizes the number of injections recovered at a FAR threshold of one per two months; for the NSBH scans, exact recovery ties are resolved in favor of larger $\nu_{\rm eff}$ and then smaller $\alpha$. The L1-calibrated parameters are held fixed within each search category and trigger-rate class and are applied to both H1 and L1.

\subsection{\label{sec:far}False alarm rate extrapolation}

The FAR of a candidate is the expected rate of noise triggers with ranking statistic larger than that of the candidate. We estimate it using the background from the same detector and analysis chunk, obtained from data without added signals after excluding time windows around catalogued events. This choice allows the FAR estimates to reflect differences in detector sensitivity and noise conditions between chunks.

For a search run trigger with ranking statistic $\ln\mathcal{L}_0$, the empirical FAR is assigned from the survival function of the background distribution,
\begin{equation}\label{eq:define_far}
\widehat{F}_{\rm emp}(\ln \mathcal{L}_0) =
\frac{N_{\rm bg}(\ln\mathcal{L}>\ln\mathcal{L}_0)}{T_{\rm bg}},
\end{equation}
where $N_{\rm bg}(\ln\mathcal{L}>\ln\mathcal{L}_0)$ is the number of background triggers louder than the candidate and $T_{\rm bg}$ is the analyzed background livetime.

The finite background livetime limits the smallest nonzero empirical FAR to $1/T_{\rm bg}$. For these week-scale backgrounds, this counting scale is of order $50\,\mathrm{yr}^{-1}$, so even the adopted detection threshold of one per two months relies on the fitted tail. To estimate FARs for candidates with ranking statistics beyond the observed background tail, we fit and extrapolate the tail of the background distribution. These extrapolated FARs depend on the fitted tail model and carry uncertainty from the finite background sample and the extrapolation, as also noted for single-detector significance estimates in GWTC-3 and GWTC-4~\cite{gwtc-3,gwtc-4}.

We fit the background tail with one straight line for BNS and two continuous straight-line segments for NSBH in the $(\ln\mathcal{L},\log_{10}\mathrm{FAR})$ plane. For NSBH, the breakpoint is chosen by minimizing the weighted residual. Both fits begin at the empirical point defined by
\begin{equation}
\widehat{F}_{\rm emp}(\ln\mathcal{L}_1)=10^{-3}\,\mathrm{Hz}.
\label{eq:emp_far}
\end{equation}
For BNS, the empirical tail has an approximately constant slope. We fit the background points with empirical FAR below $10^{-3}\,\mathrm{Hz}$ by least squares, constraining the line to pass through $(\ln\mathcal{L}_1,-3)$. The FAR mapping is
\begin{equation}\label{eq:far_extrapolation_bns}
\mathrm{FAR}(\ln \mathcal{L}) =
\begin{cases}
\widehat{F}_{\mathrm{emp}}(\ln \mathcal{L}),
& \ln \mathcal{L}\le \ln \mathcal{L}_1, \\[6pt]
10^{a_1 \ln \mathcal{L}+b_1},
& \ln \mathcal{L} > \ln \mathcal{L}_1 .
\end{cases}
\end{equation}

For NSBH, the background tail changes slope at larger ranking statistics. We fit two segments jointly, requiring the fit to pass through the same lower anchor and to be continuous at the breakpoint $\ln\mathcal{L}_2$. The slopes satisfy $a_1\leq a_2\leq0$, so the second segment is no steeper than the first. The FAR mapping is
\begin{equation}\label{eq:far_extrapolation_nsbh}
\mathrm{FAR}(\ln \mathcal{L}) =
\begin{cases}
\widehat{F}_{\mathrm{emp}}(\ln \mathcal{L}),
& \ln \mathcal{L}\le \ln \mathcal{L}_1, \\[6pt]
10^{a_1 \ln \mathcal{L}+b_1},
& \ln \mathcal{L}_1< \ln \mathcal{L} \le \ln \mathcal{L}_2, \\[6pt]
10^{a_2 \ln \mathcal{L}+b_2},
& \ln \mathcal{L}>\ln \mathcal{L}_2 .
\end{cases}
\end{equation}

To choose the NSBH breakpoint, we consider trial values of $F_2$ from $10^{-6}$ to $10^{-4}\,\mathrm{Hz}$, with step sizes of $10^{-6}\,\mathrm{Hz}$ and $10^{-5}\,\mathrm{Hz}$ below and above $10^{-5}\,\mathrm{Hz}$, respectively. For each $F_2$, we set $\ln\mathcal{L}_2$ to the ranking statistic of the empirical background point closest to $F_2$ in $\log_{10}\mathrm{FAR}$. The candidate specifies the breakpoint location; the fitted FAR there is determined by the joint fit.

The empirical tail contains many more points near $10^{-3}\,\mathrm{Hz}$ than at lower FAR. We omit the highest-ranked point from the fit only; it remains in the empirical background and cumulative counts.
For both BNS and NSBH, we divide the tail at or above $\ln\mathcal{L}_1$ into ten equal-width intervals in $\log_{10}\widehat{F}_{\rm emp}$. The BNS interval boundaries are defined before this exclusion, whereas the NSBH boundaries and breakpoint candidates are determined from the retained points. Each nonempty interval receives the same total weight, divided equally among its retained points.
For each candidate breakpoint, we minimize the weighted sum of squared differences between the empirical and fitted log FARs. We retain the candidate with the smallest residual subject to the slope constraints.

The results below use offline, same-chunk backgrounds. For low-latency operation, each detector's background could instead be estimated from the preceding seven days and updated daily. Its trigger-rate class could be updated at the same time using the rate over that seven-day window, normalized by the historical median for the same detector and search category. New triggers would then use the most recent background fit, without using data recorded after the candidate.

\subsection{\label{sec:test_methods}Validation methods}

We assess the ranking statistic and FAR assignment described above through three tests. First, we compare cumulative background counts from O3a BNS and O3b NSBH searches in H1 and L1 with the expectations for their analyzed livetimes. Second, we estimate sensitive distances using simulated injections from the GWTC-3 sensitivity study and compare them with the corresponding GWTC-3 results. Third, we test the recovery of GW190425, GW200105\_162426 and GW230529\_181500.

\subsubsection{Consistency of false alarm rate assignment}

We assess the FAR estimates using the non-injection searches over the entire O3a run for BNS and the entire O3b run for NSBH. For each detector and chunk, we evaluate the background triggers using the parameters assigned to its search category and trigger-rate class. We then fit the background tail and assign FARs using that chunk's own background, after excluding windows of $\pm1\,\mathrm{s}$ around catalogued events. We then combine the triggers from all chunks, retaining the FAR estimated for each trigger in its own chunk. The cumulative counts are formed separately for each detector and source category.
Because each chunk is used for both fitting and counting, this is an internal consistency check rather than a held-out validation. The Poisson bands are pointwise reference intervals and do not account for correlations between retained triggers.

For a FAR threshold $F_{\rm th}$, the expected number of background triggers is
\begin{equation}
N_{\rm exp}(F_{\rm th})=T_{\rm val}F_{\rm th},
\end{equation}
where $T_{\rm val}=\sum_c T_c$ is the total analyzed livetime of the chunks included in the corresponding cumulative count, after the event windows have been removed. Equivalently, for an inverse FAR threshold
$\tau=1/F_{\rm th}$,
\begin{equation}
N_{\rm exp}(\tau)=\frac{T_{\rm val}}{\tau}.
\end{equation}
We compare the number of triggers with $\mathrm{IFAR}\geq\tau$ with this expectation to assess the agreement between the estimated FARs and the observed background counts over the full analyzed observing period.

\subsubsection{Detection efficiency}

We quantify the detection efficiency using simulated BNS and NSBH injections from the GWTC-3 sensitivity study and public sensitivity data release~\cite{gwtc-3,sensitivity_estimate}. The injection populations follow the astrophysical distributions and are summarized in Tables~\ref{tab:bns_injection_priors} and~\ref{tab:nsbh_injection_priors}. Within this release, GstLAL is the only pipeline for which corresponding single-detector BNS and NSBH sensitivity estimates are provided. Its single-detector ranking includes an empirical penalty~\cite{gstlal_single_2}. To make a fair comparison, we use the same selected O3 single-detector segments and exclude injections outside the SPIIR detector-frame mass range for both pipelines. No spin cut is applied to the injections; in particular, the BNS spin magnitudes extend to $0.4$ despite the bank limit of $0.05$.

For the BNS search, we use O3a injections in chunks 6, 13, 14, and 20, corresponding to 2.3 days of H1-only data and 2.2 days of L1-only data. These segments cover 37.3\% and 34.3\% of the O3a H1 and L1 single-detector livetimes, respectively. For the NSBH search, we use O3b injections in chunks 26, 28, and 32, corresponding to 1.5 days of H1-only data and 1.4 days of L1-only data, or 32.1\% and 30.9\% of the corresponding O3b single-detector livetimes.

We associate each injection with the highest-SNR trigger within \(\pm1.1\,\mathrm{s}\) and use that trigger's FAR. An injection is considered recovered at a FAR threshold \(\mathcal{F}\) if the associated trigger is assigned \(\mathrm{FAR}\le \mathcal{F}\). We report the sensitivity in terms of the sensitive distance,
\begin{equation}
D_{\rm sen}=\left(\frac{3V_{\rm sen}}{4\pi}\right)^{1/3},
\end{equation}
where \(V_{\rm sen}\) is evaluated using a weighted Monte Carlo estimator~\cite{pycbc_open} and its finite-injection variance~\cite{farr_selection_2019}, as given in Eqs.~(6) and (7) of Ref.~\cite{sens_d_3}. The same single-detector time selection and detector-frame mass-support indicator are applied to both pipelines. Because the detector-frame support depends on redshift, the target joint mass prior is renormalized within the template range at each redshift, while the time-conditioned value of \(N_{\rm draw}\) is left unchanged after the support cut. The analytic variance is propagated through the transformation above to obtain the asymmetric \(1\sigma\) uncertainty on \(D_{\rm sen}\). 

\begin{table}[!t]
    \caption{BNS injection parameter priors and ranges for the GWTC-3 sensitivity study. Mass ranges are source-frame. Spin magnitudes are uniform over the listed ranges, with isotropic directions.\label{tab:bns_injection_priors}}
    \begin{ruledtabular}
    \begin{tabular}{lll}
        Parameter & Prior & Range \\
        \hline
        $m_1\ (M_\odot)$ & $p(m_1)\propto(m_1-1\,M_\odot)$ & $[1,\,2.5]$ \\
        $m_2\ (M_\odot)$ & Uniform & $[1,\,m_1]$ \\
        $a_1$ & Uniform; isotropic & $[0,\,0.4]$ \\
        $a_2$ & Uniform; isotropic & $[0,\,0.4]$ \\
        Redshift & $\propto (dV_c/dz)/(1+z)$ & $[0,\,0.15]$ \\
        Right ascension & Uniform & $[0,\,2\pi]$ \\
        Declination & Cosine & $[-\pi/2,\,\pi/2]$ \\
        Inclination & Sine & $[0,\,\pi]$ \\
        Polarization & Uniform & $[0,\,2\pi]$ \\
        Network SNR & Computational cut & $\geq 6$ \\
    \end{tabular}
    \end{ruledtabular}
\end{table}

\begin{table}[!t]
    \caption{NSBH injection parameter priors and ranges for the GWTC-3 sensitivity study. Mass ranges are source-frame. Spin magnitudes are uniform over the listed ranges, with isotropic directions.\label{tab:nsbh_injection_priors}}
    \begin{ruledtabular}
    \begin{tabular}{lcc}
        Parameter & Prior & Range \\
        \hline
        $m_1\ (M_\odot)$ & $p(m_1)\propto m_1^{-2.35}$ & $[2.5,\,60]$ \\
        $m_2\ (M_\odot)$ & Uniform & $[1,\,2.5]$ \\
        $a_1$ & Uniform; isotropic & $[0,\,0.998]$ \\
        $a_2$ & Uniform; isotropic & $[0,\,0.4]$ \\
        Redshift & $\propto (dV_c/dz)/(1+z)$ & $[0,\,0.25]$ \\
        Right ascension & Uniform & $[0,\,2\pi]$ \\
        Declination & Cosine & $[-\pi/2,\,\pi/2]$ \\
        Inclination & Sine & $[0,\,\pi]$ \\
        Polarization & Uniform & $[0,\,2\pi]$ \\
        Network SNR & Computational cut & $\geq 6$ \\
    \end{tabular}
    \end{ruledtabular}
\end{table}

\subsubsection{\label{sec:revisit_methods}Recovery tests for known single-detector events}

We further test the method on three BNS and NSBH events whose catalog significance was determined from a single detector in GWTC-3 and GWTC-4: GW190425~\cite{gw190425}, GW200105\_162426~\cite{gw200105}, and GW230529\_181500~\cite{gw230529,gwtc-4}. L1 and V1 were operating during the first two events, but their reported significance was determined by the Livingston trigger~\cite{gwtc-2.1,gwtc-3}. GW230529\_181500 occurred during O4a when only L1 provided usable data~\cite{gw230529}. Accordingly, all SNRs quoted in this comparison are Livingston single-detector values.

For the O3 events, we estimate FARs using the L1 background from the chunk containing each event, excluding windows of $\pm1\,\mathrm{s}$ around catalogued events, as described in Sec.~\ref{sec:far}. For GW230529\_181500, we use L1 trigger quantities and a separate two-segment tail fit to the available background files from the first five pre-event one-day windows of O4a chunk~1.
For the chunk~3 BNS and chunk~31 high-rate NSBH calibrations, the non-injection backgrounds exclude $\pm1\,\mathrm{s}$ around catalogued events; the chunk~31 injection overlapping GW200105\_162426 is also excluded from the calibration objective.

\section{\label{sec:results}Results}

We evaluate the method described in Sec.~\ref{sec:methods} through three tests: the consistency of FAR estimates with the cumulative background counts over the O3a BNS and O3b NSBH searches, the sensitive distances estimated from simulated injections, and the recovery of three previously reported single-detector events.

\subsection{Parameter calibration}
\label{sec:calibration_results}

\begin{figure*}[!t]
  \centering
  \includegraphics[width=\textwidth]{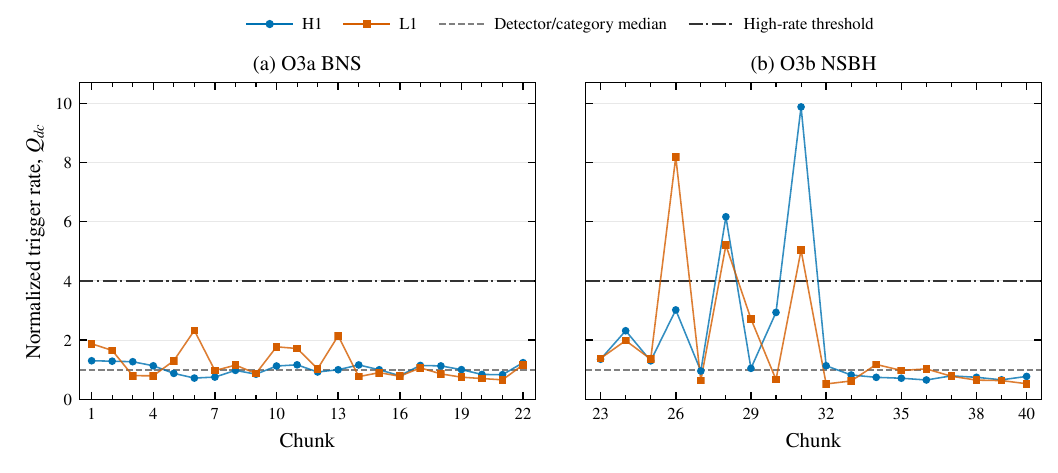}
  \caption{Normalized rates of non-injection triggers with $\rho_m\geq8$ and $\chi_r^2>3$ in (a) the O3a BNS and (b) the O3b NSBH searches. For each detector and search category, $Q_{dc}$ is the rate in chunk $c$ divided by the median over the corresponding observing period. The dashed line marks the median, and the dash-dotted line marks the high-rate threshold, $Q_{dc}=4$. Windows of $\pm1\,\mathrm{s}$ around catalogued events are excluded from the counts and livetime. }
  \label{fig:chunk_trigger_rates}
\end{figure*}

Figure~\ref{fig:chunk_trigger_rates} shows the variation across chunks of the trigger-rate ratio $Q_{dc}$ defined in Eq.~\eqref{eq:chunk_rate_ratio}. Applying the same classification criterion to both searches, all O3a BNS chunks in H1 and L1 satisfy $Q_{dc}\leq4$ and belong to the nominal-rate class. In the O3b NSBH search, chunks~28 and~31 in H1 and chunks~26, 28, and~31 in L1 exceed the high-rate threshold.
The calibration samples are L1 injections from chunk~3 for BNS and from chunks~23 and~31 for nominal-rate and high-rate NSBH, respectively. %

We adopt $(\alpha,\nu_{\rm eff})=(0.8,220)$ for BNS. For NSBH, the adopted values are $(0.2,490)$ for nominal-rate chunks and $(0.1,200)$ for high-rate chunks. These choices recover 902, 334, and 471 injections in the respective L1 calibration samples at the adopted FAR threshold.
For high-rate NSBH chunks, the adopted value $\alpha=0.1$ is the lowest value scanned.

\subsection{Consistency of false alarm rate assignment}\label{sec:far_extrapolation_results}

\begin{figure*}[!t]
  \centering
  \begin{tabular}{cc}
    \includegraphics[width=0.46\textwidth]{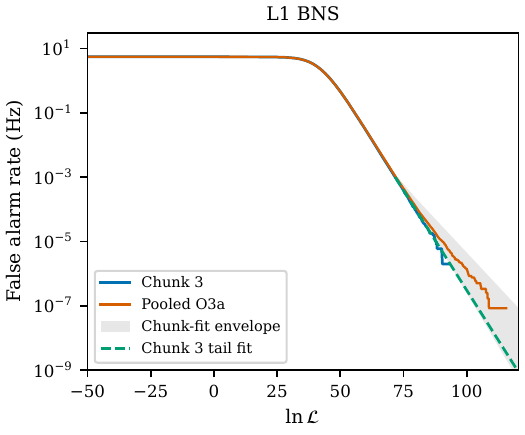} &
    \includegraphics[width=0.46\textwidth]{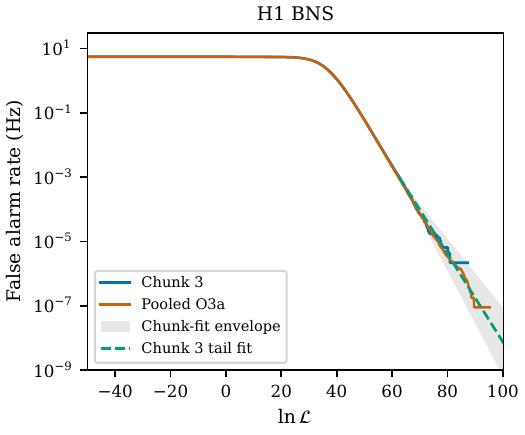} \\
    \small (a) & \small (b) \\[6pt]
    \includegraphics[width=0.46\textwidth]{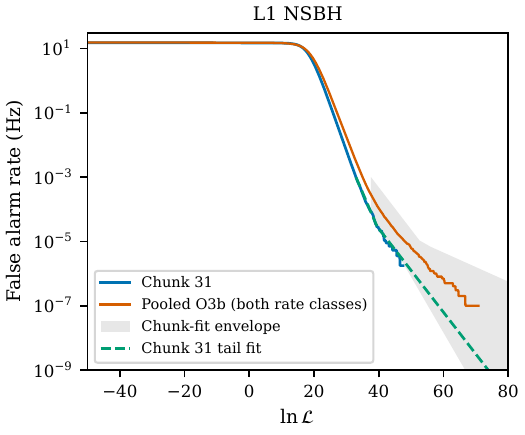} &
    \includegraphics[width=0.46\textwidth]{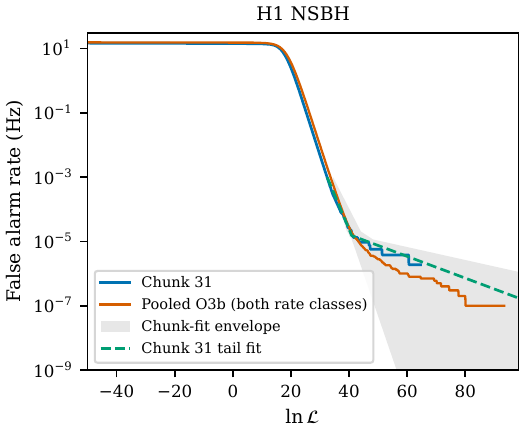} \\
    \small (c) & \small (d)
  \end{tabular}
    \caption{False alarm rate (FAR) tail fits for (a) L1 BNS, (b) H1 BNS, (c) L1 NSBH, and (d) H1 NSBH. Blue curves show chunk~3 for BNS and chunk~31 for NSBH. Orange curves pool the O3a BNS or O3b NSBH background across all chunks of each detector. The pooled rate is the cumulative trigger count divided by the summed livetime. Windows of $\pm1\,\mathrm{s}$ around catalogued events are excluded. Green dashed curves show the single-linear BNS and continuous broken-linear NSBH fits to the reference chunks, as described in Sec.~\ref{sec:far}. Shaded regions show the envelope of all individual-chunk tail fits in each panel.}
  \label{fig:far_extrapolation}
\end{figure*}

Figure~\ref{fig:far_extrapolation} illustrates the background-tail models using chunk~3 for BNS and chunk~31 for NSBH in both detectors. The comparison curves pool the O3a BNS and O3b NSBH backgrounds across all chunks separately for each detector, with both NSBH trigger-rate classes included. These pooled curves summarize the observed backgrounds rather than define a common FAR mapping.
The shaded envelopes show the chunk-to-chunk variation of the tail mapping; each chunk still uses its own background for FAR assignment.

The BNS tail is modeled by a single line in the $(\ln\mathcal{L},\log_{10}\mathrm{FAR})$ plane. The chunk~3 fits have weighted $R^2=0.9964$ for L1 and $0.9738$ for H1, using the equal-total-weight logarithmic FAR intervals defined in Sec.~\ref{sec:far}. For L1, the pooled O3a tail lies above the reference fit at low FAR but remains within the envelope of the 22 individual-chunk fits over their common plotted tail range. The widening envelopes illustrate the variation between chunks.

Compared to BNS, the NSBH background develops a shallower terminal tail. For L1 chunk~31, the jointly fitted slopes are $-0.238$ and $-0.130$, with weighted $R^2=0.9974$; the corresponding H1 fit has slopes $-0.218$ and $-0.034$, with weighted $R^2=0.9904$. The envelopes include both trigger-rate classes and widen toward lower FAR, where fewer empirical tail points constrain the extrapolation and the fitted terminal slope varies between chunks.

To investigate the nature of the shallower NSBH background tail, we examine the 16 L1 background triggers in chunk~31 above the fitted breakpoint, $\ln\mathcal{L}_2=39.33$. Grouping consecutive triggers separated by at most $10\,\mathrm{s}$ yields 15 groups, each represented by its highest-ranked trigger. Of these representatives, 13 coincide within $\pm5\,\mathrm{s}$ with entries in the Gravity Spy catalogue~\cite{gravity_spy_o3} whose classification confidence exceeds 0.5. The associated glitch classes are listed in Table~\ref{tab:nsbh_gravity_spy}. These temporal associations suggest a contribution from several transient classes.

\begin{table}[!t]
    \caption{Gravity Spy glitch classes associated with the chunk~31 terminal-tail SPIIR time groups. Consecutive SPIIR triggers separated by at most \(10\,\mathrm{s}\) are combined, and each group is represented by its largest-\(\ln\mathcal{L}\) trigger. Each representative contributes at most once: the table reports the closest catalogue peak within \(\pm5\,\mathrm{s}\) whose glitch-classification confidence is greater than \(0.5\).}
    \label{tab:nsbh_gravity_spy}
    \begin{ruledtabular}
    \begin{tabular}{lc}
    Gravity Spy class & Matched SPIIR time groups \\
    \hline
    \texttt{Extremely\_Loud}       & 2 \\
    \texttt{Fast\_Scattering}      & 4 \\
    \texttt{Low\_Frequency\_Burst} & 1 \\
    \texttt{Power\_Line}           & 1 \\
    \texttt{Repeating\_Blips}      & 1 \\
    \texttt{Scattered\_Light}      & 4 \\
    \hline
    Total                            & 13
    \end{tabular}
    \end{ruledtabular}
\end{table}

\begin{figure*}[!t]
  \centering
  \begin{tabular}{@{}cc@{}}
    \includegraphics[width=0.48\textwidth]{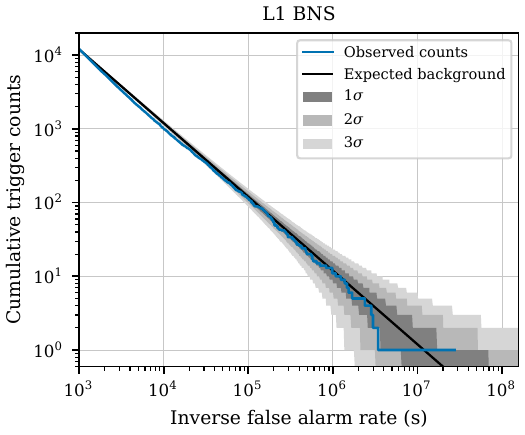} &
    \includegraphics[width=0.48\textwidth]{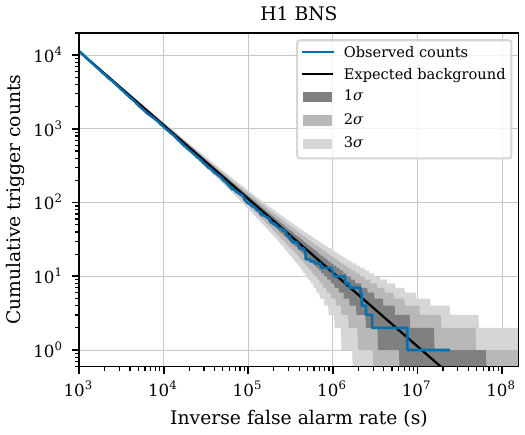} \\
    \small (a) & \small (b) \\[0.7ex]
    \includegraphics[width=0.48\textwidth]{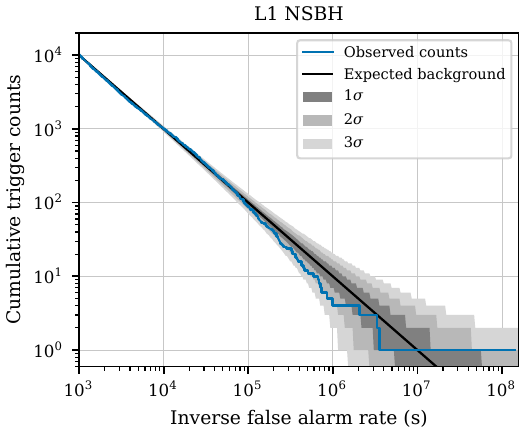} &
    \includegraphics[width=0.48\textwidth]{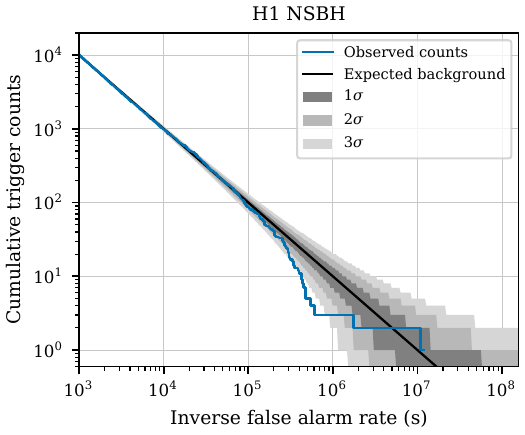} \\
    \small (c) & \small (d)
  \end{tabular}
  \caption{Cumulative background-trigger counts as a function of inverse false alarm rate (IFAR), in seconds. Panels (a)--(d) show the L1 BNS, H1 BNS, L1 NSBH, and H1 NSBH results, respectively. Blue curves show the observed counts, black curves show the expected background counts, and shaded bands indicate pointwise $1\sigma$, $2\sigma$, and $3\sigma$ Poisson reference intervals.}
  \label{fig:ifar_counts}
\end{figure*}

Figure~\ref{fig:ifar_counts}, panels (a)--(d), shows the L1 BNS, H1 BNS, L1 NSBH, and H1 NSBH cumulative background-trigger counts, respectively.

At a FAR threshold of one per two months, the L1 BNS, H1 BNS, L1 NSBH, and H1 NSBH searches contain 1, 2, 1, and 2 background triggers, compared with expectations of 2.31, 2.17, 1.92, and 1.93, respectively. None of the four searches shows an excess above the upper edge of its pointwise $3\sigma$ Poisson reference band over the plotted range.
Both NSBH searches show deficits below the lower edge over part of the range, consistent with conservative FAR assignment there.

\subsection{\label{sec:sensitivity}Detection efficiency}

\begin{figure*}[!t]
  \centering
  \begin{tabular}{cc}
    \includegraphics[width=0.46\textwidth]{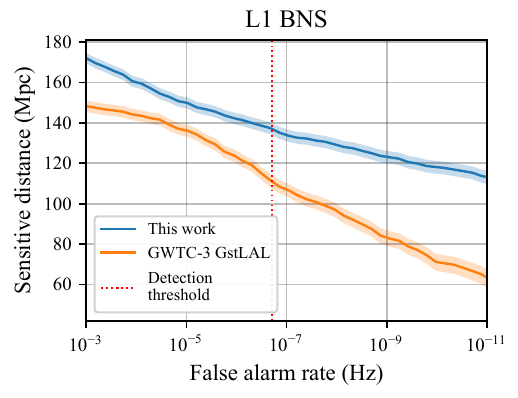} &
    \includegraphics[width=0.46\textwidth]{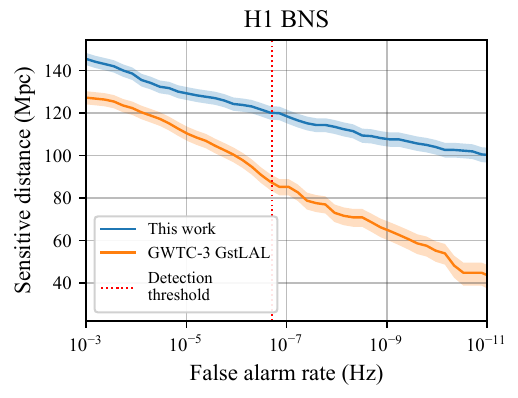} \\
    \small (a) & \small (b) \\[0.4ex]
    \includegraphics[width=0.46\textwidth]{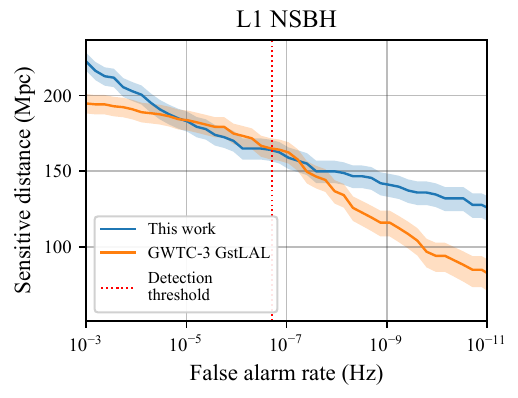} &
    \includegraphics[width=0.46\textwidth]{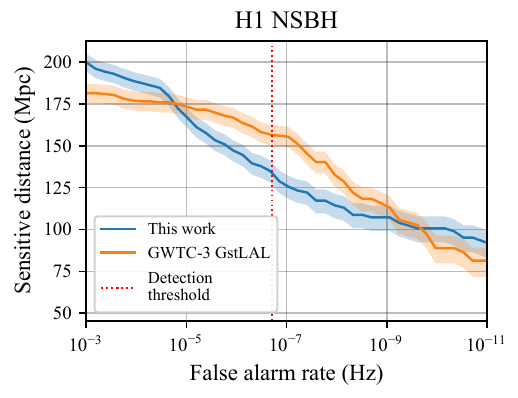} \\
    \small (c) & \small (d)
  \end{tabular}
  \caption{Sensitive distance as a function of FAR threshold for the BNS and NSBH searches using O3 data and simulated injections. Panels (a)--(d) show L1 BNS, H1 BNS, L1 NSBH, and H1 NSBH, respectively. Blue curves show the SPIIR search using the ranking statistic developed here; orange curves show GstLAL sensitivities evaluated from the GWTC-3 sensitivity release over the same selected data segments and the same injection population conditioned on the SPIIR detector-frame mass range. The shaded bands show the propagated \(1\sigma\) finite-injection Monte Carlo uncertainty on \(D_{\rm sen}\), and the vertical dashed line marks the detection threshold of one per two months.}
  \label{fig:sensitive_distance}
\end{figure*}

Figure~\ref{fig:sensitive_distance} shows the resulting sensitive distance as a function of FAR threshold for the BNS and NSBH searches in L1 and H1. %
The selected single-detector segments contain too few injections for statistically useful binned estimates, so we report BNS and NSBH sensitive distances over the selected detector-frame mass range.

In Fig.~\ref{fig:sensitive_distance}, at the detection threshold of one false alarm per two months, the SPIIR BNS sensitive distances are $136.5^{+2.9}_{-3.0}$~Mpc for L1 and $120.2^{+3.1}_{-3.3}$~Mpc for H1, compared with $110.5^{+3.2}_{-3.4}$~Mpc and $87.2^{+3.6}_{-3.9}$~Mpc for GstLAL, respectively. The quoted uncertainties are propagated $1\sigma$ finite-injection Monte Carlo uncertainties. These central estimates correspond to increases of approximately 24\% and 38\% in sensitive distance, or factors of 1.9 and 2.6 in sensitive volume.

For the NSBH search, the SPIIR sensitive distances at the same threshold are $164.1^{+6.7}_{-7.3}$~Mpc for L1 and $134.9^{+6.3}_{-7.0}$~Mpc for H1, compared with $165.1^{+6.7}_{-7.3}$~Mpc and $156.0^{+5.9}_{-6.4}$~Mpc for GstLAL, respectively. The central estimates are less than 1\% and approximately 14\% lower, respectively. The L1 sample includes high-rate chunks~26 and~28 and nominal-rate chunk~32; the H1 sample also includes high-rate chunk~28. The L1 estimates have overlapping individual finite-injection uncertainty intervals. In the low-FAR regime, the sensitivity depends on the sparsely sampled background tail. A shallower fitted terminal slope assigns higher FARs to high-ranking triggers and can reduce injection recovery.

\begin{figure*}[!t]
  \centering
  \begin{tabular}{cc}
    \includegraphics[width=0.46\textwidth]{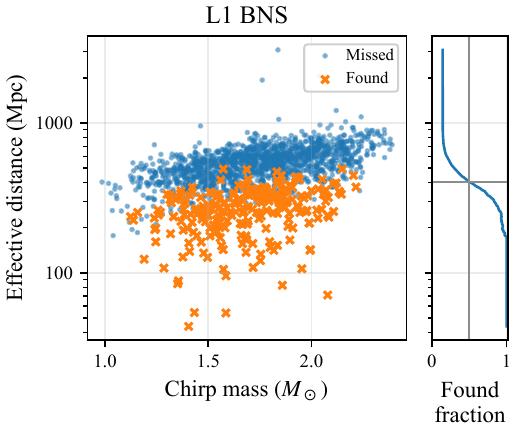} &
    \includegraphics[width=0.46\textwidth]{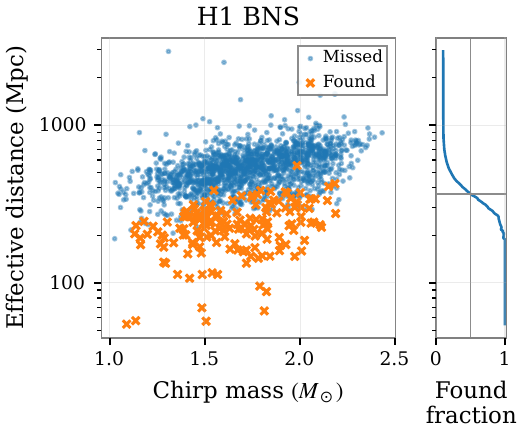} \\
    \small (a) & \small (b) \\[0.4ex]
    \includegraphics[width=0.46\textwidth]{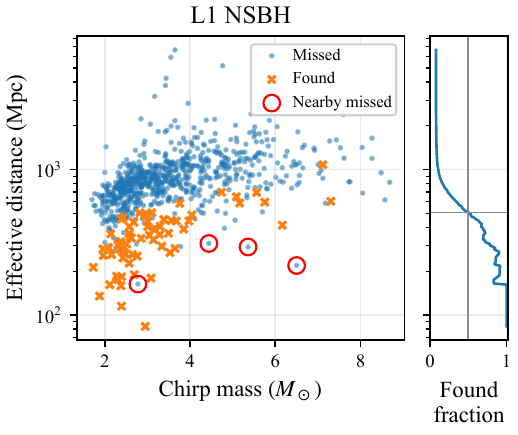} &
    \includegraphics[width=0.46\textwidth]{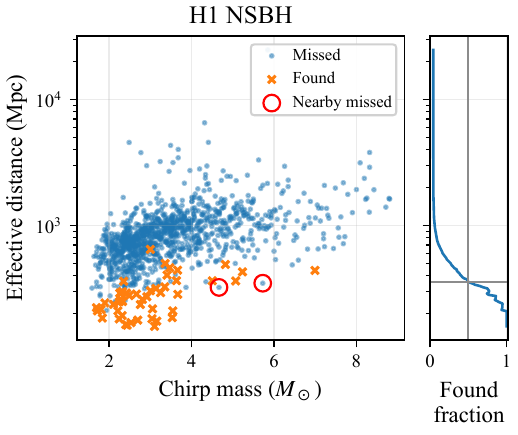} \\
    \small (c) & \small (d)
  \end{tabular}
  \caption{Missed and found injections at the threshold of one false alarm per two months. Panels (a)--(d) show L1 BNS, H1 BNS, L1 NSBH, and H1 NSBH, respectively. Orange crosses denote found injections and blue points denote missed injections. Red open circles mark the nearby missed injections examined in Appendix~\ref{app:nearby_missed}. The side panels show the fraction of these matched injections that were found at or below a given effective distance.}
  \label{fig:found_missed}
\end{figure*}

Fig.~\ref{fig:found_missed} shows the distribution of missed and found injections over chirp mass and effective distance at the detection threshold of one per two months. The effective distance, $d_{\rm eff}$, is the distance at which an optimally located and oriented binary with the same intrinsic parameters would produce the same signal amplitude in the detector. The recovered population is concentrated toward smaller effective distances, as expected from the distance scaling of the signal amplitude. The recovery boundary also depends on chirp mass, with the more massive systems remaining detectable at larger effective distances. The side panels show the cumulative fraction of matched injections that are found at or below a given effective distance,
\begin{equation}
  f_{\rm found}(\le d)=
  \frac{N_{\rm found}(d_{\rm eff}\le d)}
       {N_{\rm matched}(d_{\rm eff}\le d)} ,
\end{equation}
where \(N_{\rm matched}\) is the number of injections with at least one corresponding SPIIR trigger within the \(\pm1.1\,\mathrm{s}\) matching window, and \(N_{\rm found}\) is the subset whose assigned FAR is at or below the detection threshold. This cumulative fraction equals 0.5 at effective distances of 405~Mpc for L1 BNS, 366~Mpc for H1 BNS, 507~Mpc for L1 NSBH, and 357~Mpc for H1 NSBH.

Six highlighted missed NSBH injections, marked by red open circles, occur at relatively small effective distances. Appendix~\ref{app:nearby_missed} examines their matched-trigger properties and the corresponding strain Q-transform maps. Local transient noise can increase the signal-consistency penalty and reduce the ranking statistic. The appendix also repeats the L1 NSBH recovery test on another data chunk with nominal trigger rate, shown in Fig.~\ref{fig:ck25_nsbh_missed_found}, as a supplementary recovery check. In that check, the missed and found injections show the expected separation in effective distance, with no comparable concentration of nearby missed injections.

\subsection{\label{sec:revisit}Recovery tests for known single-detector events}

We recovered all three BNS and NSBH events whose catalog significance was determined from a single detector in GWTC-3 and GWTC-4 at the detection threshold of one per two months. Table~\ref{tab:known_events} compares our recovered triggers with the published results.

GW190425 is recovered with an L1 SNR of 12.3 and an extrapolated FAR below $1.0\times10^{-5}\,\mathrm{yr}^{-1}$. GW200105\_162426 is recovered with an L1 SNR of 13.6 and an extrapolated FAR of $3.36\times10^{-3}\,\mathrm{yr}^{-1}$.
GW230529\_181500 is recovered with an L1 SNR of 11.2 and an extrapolated FAR below $1.0\times10^{-5}\,\mathrm{yr}^{-1}$. For all events, the background is drawn from the same detector and chunk as the event, with the event excluded from the fit. All satisfy the adopted detection threshold of one per two months.

\begin{table*}[p]
    \caption{Recovery tests for BNS and NSBH events with single-detector significance in GWTC-3 and GWTC-4. Catalog values are from GWTC-3 GstLAL for GW190425 and GW200105\_162426, and from GWTC-4 GstLAL, MBTA, and PyCBC for GW230529\_181500~\cite{gwtc-2.1,gwtc-3,gwtc-4}. Detector columns list available data; all SNRs here are L1-only. SPIIR uses L1 backgrounds from O3 chunk~3, ~31, O4 chunk~1 for GW190425, GW200105\_162426, and GW230529\_181500, respectively. SPIIR FARs below the empirical counting scale of the chunk background, $1/T_{\rm bg}$, are extrapolated from the fitted tail. The reporting floor is $1.0\times10^{-5}$ yr$^{-1}$; smaller fitted FARs are shown with a ``$<$'' sign. FARs from different pipelines use different backgrounds and methods and are not directly comparable.\label{tab:known_events}}
    \begin{ruledtabular}
    \begin{tabular}{lcccccccc}
        GW Event & Source &
        \multicolumn{4}{c}{GWTC} &
        \multicolumn{3}{c}{SPIIR single-detector} \\
        \cline{3-6} \cline{7-9}
         & &
        Pipeline & FAR (yr$^{-1}$) & SNR & Detector &
        FAR (yr$^{-1}$) & SNR & Detector \\
        \hline
        GW190425 & BNS
        & GstLAL & 0.034 & 12.9 & L1V1
        & $<1.0\times10^{-5}$ & 12.3 & L1 \\
        GW200105\_162426 & NSBH
        & GstLAL & 0.2 & 13.6 & L1V1
        & $3.36\times10^{-3}$ & 13.6 & L1 \\
        \multirow{3}{*}{GW230529\_181500} & \multirow{3}{*}{NSBH}
        & GstLAL & $5.8\times10^{-3}$ & 11.8 & L1
        & \multirow{3}{*}{$<1.0\times10^{-5}$} & \multirow{3}{*}{11.2} & \multirow{3}{*}{L1} \\
        & & MBTA & $2.2\times10^{-4}$ & 11.4 & L1 & & & \\
        & & PyCBC & $1.0\times10^{-3}$ & 11.7 & L1 & & & \\
    \end{tabular}
    \end{ruledtabular}
\end{table*}

\section{\label{sec:conclusion}Conclusion}

We have developed a hybrid log-likelihood ratio for single-detector BNS and NSBH searches with SPIIR. The statistic combines a signal-consistency model for non-Gaussian glitches with an SNR profile-likelihood-ratio term derived under a Gaussian-noise model.
The glitch model's parameters are calibrated separately for different search categories and trigger-rate classes to adapt to varying levels of glitch activity.
FARs are estimated from the background of the same detector and analysis chunk, with a single linear tail for BNS and a continuous broken-linear tail for NSBH in the $(\ln\mathcal{L},\log_{10}\mathrm{FAR})$ plane.

We assessed the FAR estimates by pooling background-trigger counts after assigning FARs separately within each detector and chunk. The BNS and NSBH searches show no excess above the upper edges of the pointwise $3\sigma$ Poisson reference bands over the plotted IFAR range.

We measured the detection sensitivity using simulated signals from the GWTC-3 sensitivity study and compared it with GstLAL over the same selected data segments and injection parameter ranges. At the threshold of one per two months, the SPIIR BNS sensitive distances are $136.5^{+2.9}_{-3.0}$~Mpc for L1 and $120.2^{+3.1}_{-3.3}$~Mpc for H1, compared with $110.5^{+3.2}_{-3.4}$~Mpc and $87.2^{+3.6}_{-3.9}$~Mpc for GstLAL. The central estimates are approximately 24\% and 38\% higher, respectively. For NSBH, SPIIR reaches $164.1^{+6.7}_{-7.3}$~Mpc for L1 and $134.9^{+6.3}_{-7.0}$~Mpc for H1, compared with $165.1^{+6.7}_{-7.3}$~Mpc and $156.0^{+5.9}_{-6.4}$~Mpc for GstLAL.

We further recovered GW190425, GW200105\_162426 and GW230529\_181500 using their own O3 and O4 chunks as background. All three reported event results lie below the adopted detection threshold.

The NSBH background investigation suggests that transient noise contributes to the high-ranking background tail.
Of the 15 representative background triggers above the fitted breakpoint in the chunk~31 NSBH background, 13 (87\%) are temporally associated with glitches classified by Gravity Spy. The remaining 2 are unclassified by this catalogue-matching test.
Future work will extend the framework to single-detector BBH searches, where such transients may have an even stronger effect. Information beyond the current $(\rho_m,\chi_r^2)$ feature space, including time--frequency band occupancy and machine-learning scores from methods such as SigMA-Net~\cite{sunil_ml_1,sunil_ml_2}, could help identify and downrank transient-like triggers, suppress the terminal-tail excess, restore a more nearly linear FAR relation, and improve search sensitivity.

A coherent multi-detector extension would give the SPIIR pipeline a common ranking framework for single-detector and multi-detector operation. Incorporating coherence information between detectors may also improve the sensitivity of future online and offline searches.

\section*{Acknowledgments}
This research made use of the OzSTAR national supercomputing facility hosted by Swinburne University of Technology, which is supported by the Australian Government through the National Collaborative Research Infrastructure Strategy (NCRIS) and by the Victorian Government through the Victorian Higher Education State Investment Fund (VHESIF). This work also relied on computational resources provided by the LIGO Laboratory, supported by the U.S. National Science Foundation under Grant Nos. PHY-0757058 and PHY-0823459, and was additionally supported by the JST ASPIRE Program of Japan under Grant No. JPMJAP2320. W.~G. thanks Kipp Cannon for discussions. This work was supported by the National Research Foundation of Korea (NRF) grants funded by the Korea Ministry of Science and ICT (MSIT) (RS-2025-00555178) for JCK.

We acknowledge the Gravitational Wave Open Science Center (GWOSC), a service of the LIGO Scientific Collaboration, the Virgo Collaboration, and KAGRA, for access to public data and software~\cite{o3_open_data,o4_open_data}. This material is based upon work supported by NSF's LIGO Laboratory which is a major facility fully funded by the National Science Foundation. We thank the Gravity Spy team and volunteers for the glitch classifications used in this work~\cite{gravity_spy_o3}.

\section*{Author contributions}

The first six authors are listed in order of contribution; the remaining authors, beginning with Tim Davies, are listed alphabetically by surname.

\newpage
\appendix
\section{\label{app:nearby_missed}Nearby missed injections}

\begin{figure*}[!t]
  \centering
  \begin{tabular}{cc}
    \includegraphics[width=0.46\textwidth]{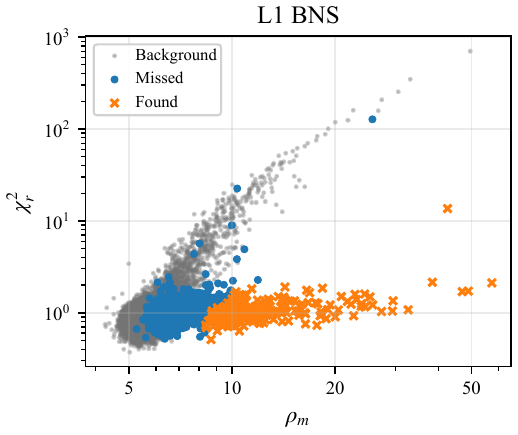} &
    \includegraphics[width=0.46\textwidth]{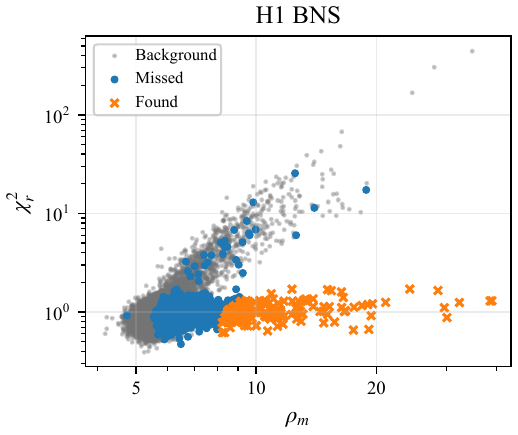} \\
    \small (a) & \small (b) \\[0.4ex]
    \includegraphics[width=0.46\textwidth]{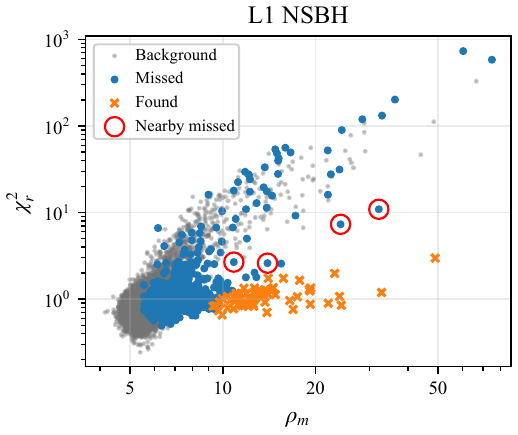} &
    \includegraphics[width=0.46\textwidth]{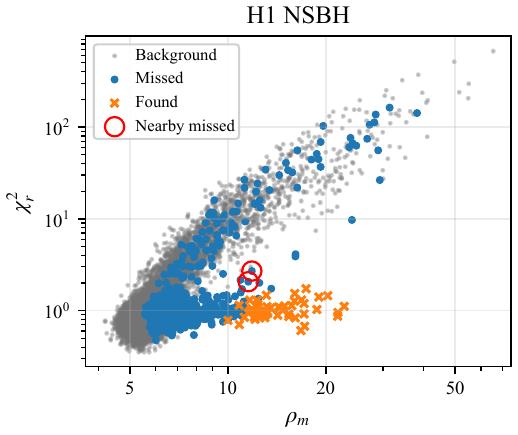} \\
    \small (c) & \small (d)
  \end{tabular}
  \caption{Simulated injections and background triggers in the $(\rho_m,\chi_r^2)$ plane. Panels (a)--(d) show L1 BNS, H1 BNS, L1 NSBH, and H1 NSBH, respectively. Gray points show background triggers. Blue points and orange crosses show the triggers associated with the missed and found injections in Fig.~\ref{fig:found_missed}, while red open circles identify the triggers associated with the nearby missed injections. Found or missed classification uses the FAR of the highest-SNR trigger within \(\pm1.1\,\mathrm{s}\) of each injection. }
  \label{fig:snr_chisq}
\end{figure*}

This appendix gives supporting checks for the six nearby missed NSBH injections marked in Fig.~\ref{fig:found_missed}. Figure~\ref{fig:snr_chisq} compares the associated triggers with the empirical background in the $(\rho_m,\chi_r^2)$ plane. Recovered injections occupy the expected region of large $\rho_m$ and $\chi_r^2$ near unity, while the highlighted missed injections are closer to the background population. These triggers are assigned FARs above the detection threshold because their SNR and signal-consistency statistic place them outside the high-ranking region.

\begin{figure*}[!t]
  \centering
  \begin{tabular}{@{}cc@{}}
    \includegraphics[width=0.48\textwidth]{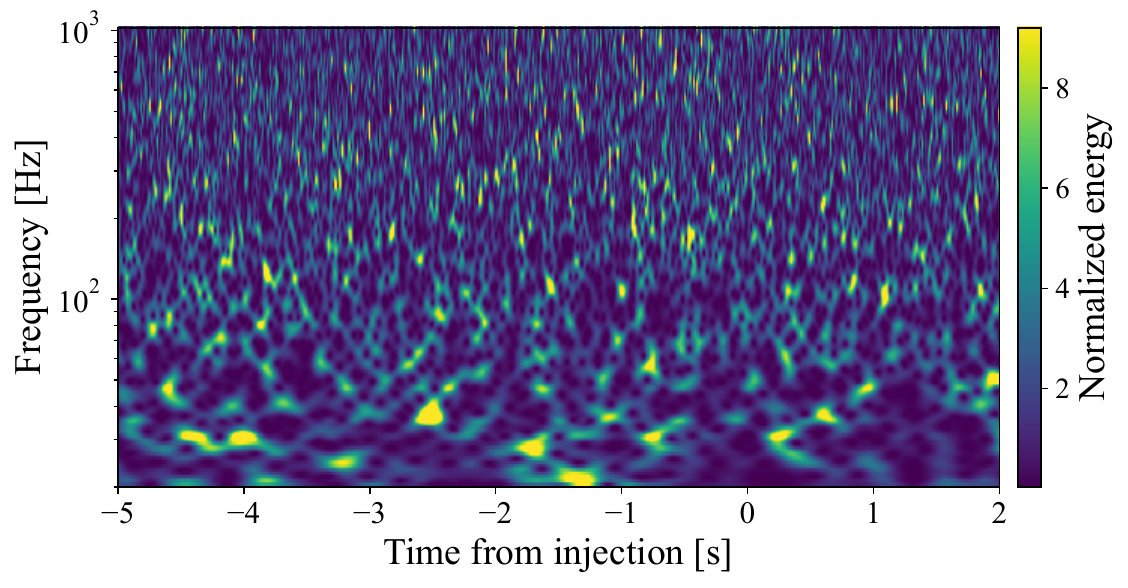} &
    \includegraphics[width=0.48\textwidth]{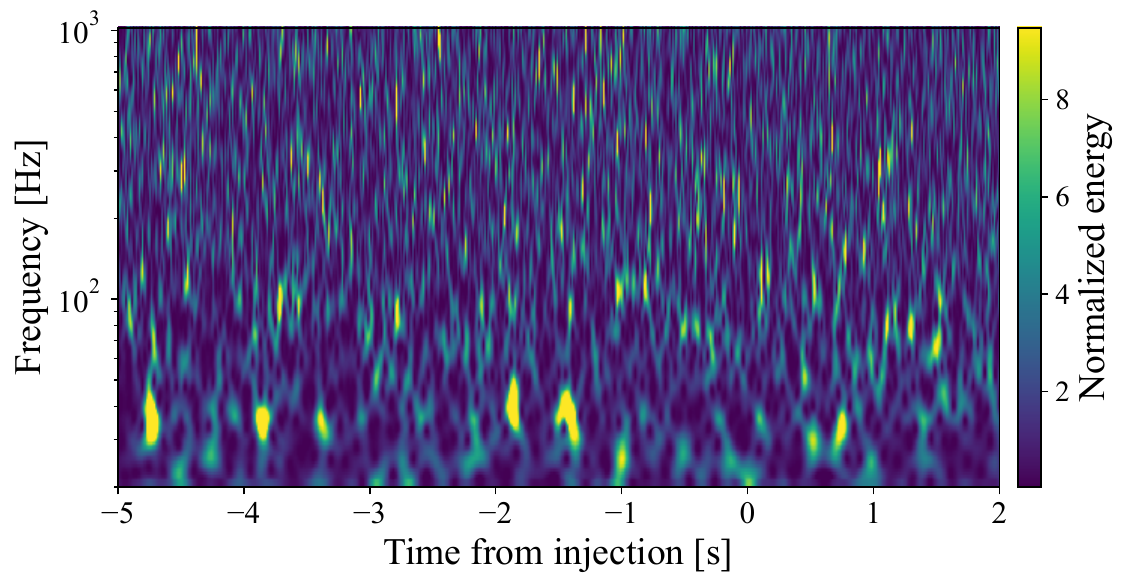} \\
    \normalsize (a) & \normalsize (b) \\[0.4ex]
    \includegraphics[width=0.48\textwidth]{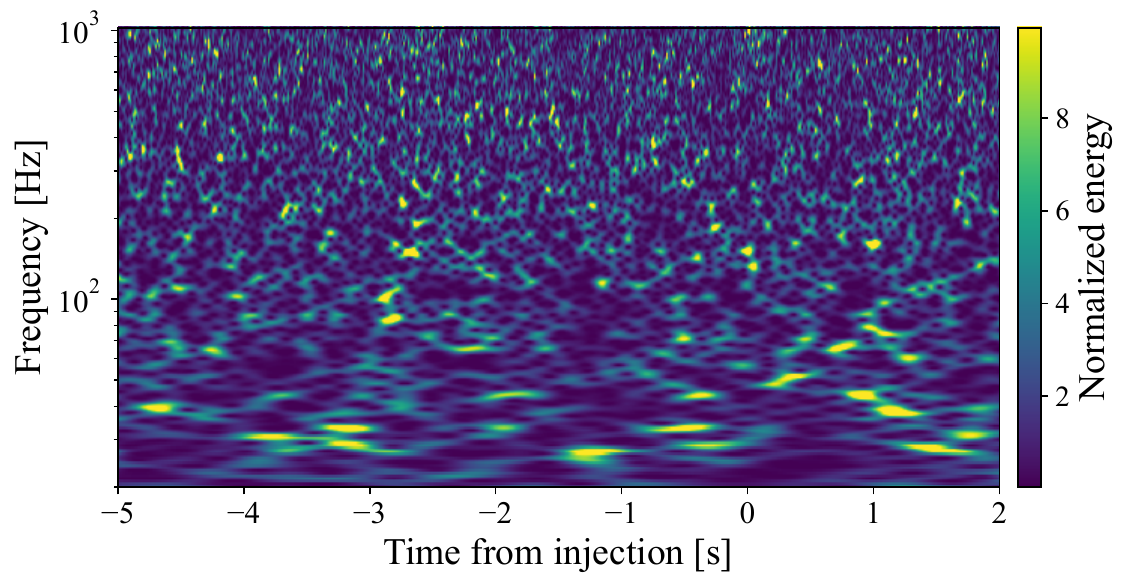} &
    \includegraphics[width=0.48\textwidth]{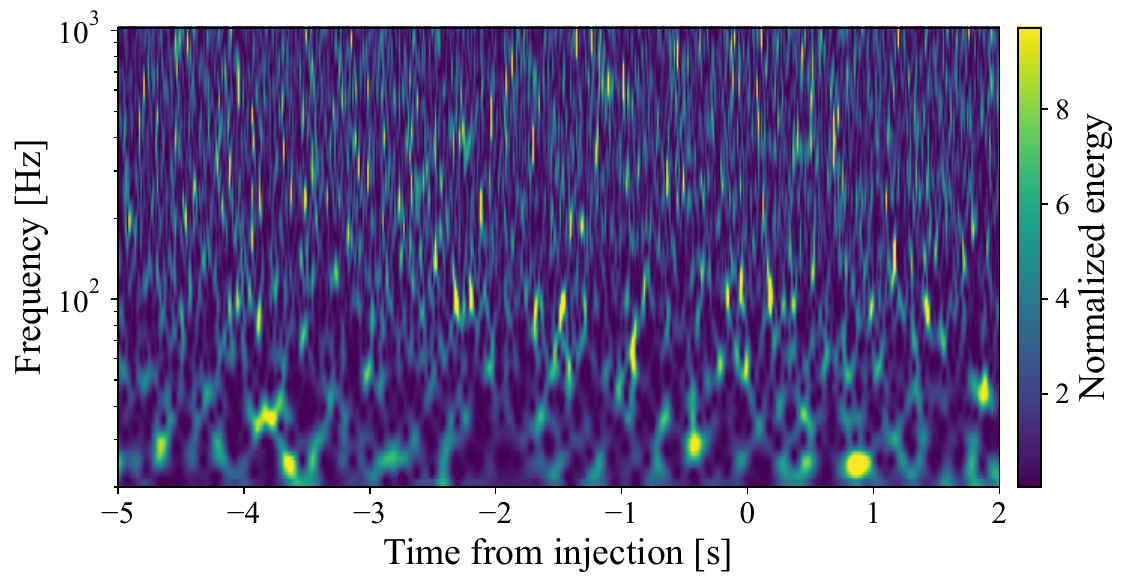} \\
    \normalsize (c) & \normalsize (d) \\[0.4ex]
    \includegraphics[width=0.48\textwidth]{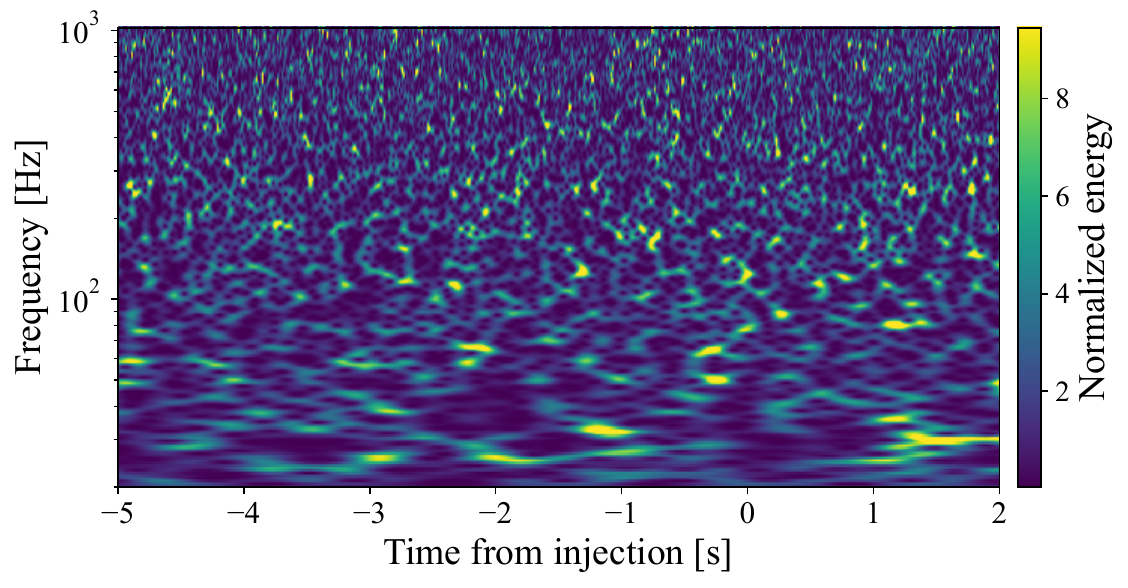} &
    \includegraphics[width=0.48\textwidth]{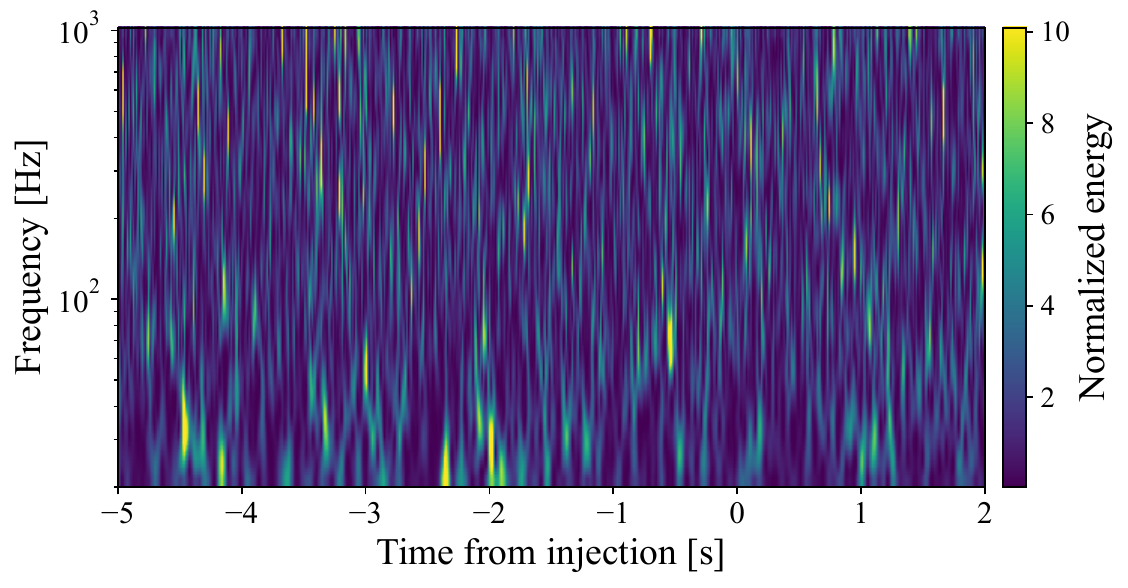} \\
    \normalsize (e) & \normalsize (f)
  \end{tabular}
  \caption{Q-transform maps of the detector strain around the six injection times highlighted in Fig.~\ref{fig:found_missed}. Panels (a)--(d) show L1 at GPS times 1259318836, 1260193204, 1258951917, and 1263143318; panels (e) and (f) show H1 at GPS times 1258990193 and 1263544186. The maps use whitening, $Q=4$--$64$, and frequencies $20$--$1024\,\mathrm{Hz}$, displaying $-5$ to $+2\,\mathrm{s}$ relative to each injection time. Simulated injection waveforms are not added to these data. }
  \label{fig:highlighted_qtransform}
\end{figure*}

Figure~\ref{fig:highlighted_qtransform} shows Q-transform maps for the six nearby missed injections. We inspect these maps qualitatively for transient structure that could contribute to the signal-consistency penalty.

We also perform an additional L1 NSBH recovery check, since this case contains the largest number of nearby missed injections in Fig.~\ref{fig:found_missed}. The check uses all L1 data from chunk~25, regardless of whether another detector was also observing, with the same injection population.

\begin{figure}[!t]
  \centering
  \includegraphics[width=\columnwidth]{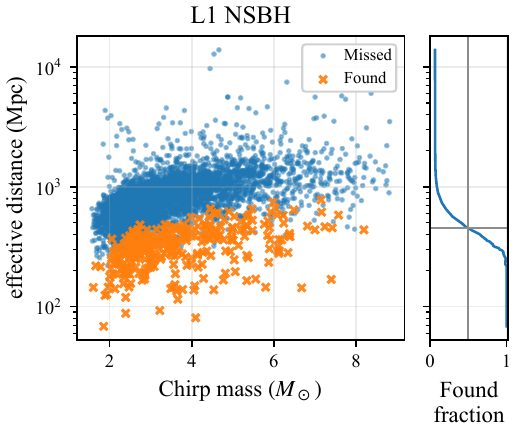}
  \caption{Additional L1 NSBH missed and found injection check using the chunk~25 data. FARs are assigned with the chunk~25 L1 NSBH background. Orange crosses denote injections recovered below the detection threshold of one per two months, and blue points denote missed injections. The side panel shows the fraction of matched injections found at or below a given L1 effective distance.}
  \label{fig:ck25_nsbh_missed_found}
\end{figure}

Figure~\ref{fig:ck25_nsbh_missed_found} shows that the recovered injections occupy the nearby region of the chirp-mass and effective-distance plane, while the missed injections are predominantly at larger effective distance. This check supports the interpretation that the nearby missed injections highlighted in Fig.~\ref{fig:found_missed} are associated with the particular data segments used in the comparison.

\clearpage

\bibliographystyle{apsrev4-2}
\bibliography{references}

\end{document}